\documentclass[conference]{IEEEtran}
\IEEEoverridecommandlockouts
\usepackage{cite}
\usepackage{amsmath,amssymb,amsfonts}
\usepackage{algorithmic}
\usepackage{graphicx}
\usepackage{textcomp}
\usepackage{xcolor}
\makeatother
\makeatletter
\usepackage{multirow, makecell}
\usepackage{array}
\usepackage{tikz}
\usepackage{tabularx}
\usepackage{booktabs}
\def\BibTeX{{\rm B\kern-.05em{\sc i\kern-.025em b}\kern-.08em
    T\kern-.1667em\lower.7ex\hbox{E}\kern-.125emX}}
\begin{document}

\title{Decoding 3D Representation of Visual Imagery EEG using Attention-based Dual-Stream Convolutional Neural Network\\
% {\footnotesize \textsuperscript{*}}
\thanks{This research was supported by the Defense Challengeable Future Technology Program of Agency for Defense Development, Republic of Korea.}
}

\author{\IEEEauthorblockN{Hyung-Ju Ahn}
\IEEEauthorblockA{\textit{Dept. Brain and Cognitive Engineering} \\
\textit{Korea University}\\
Seoul, Republic of Korea \\
hj\_ahn@korea.ac.kr}
\and
\IEEEauthorblockN{Dae-Hyeok Lee}
\IEEEauthorblockA{\textit{Dept. Brain and Cognitive Engineering} \\
\textit{Korea University}\\
Seoul, Republic of Korea \\
lee\_dh@korea.ac.kr}
}

\maketitle

\begin{abstract}
A deep neural network has been successfully applied to an electroencephalogram (EEG)-based brain-computer interface. However, in most studies, the correlation between EEG channels and inter-region relationships are not well utilized, resulting in sub-optimized spatial feature extraction. In this study, we propose an attention-based dual-stream 3D-convolutional neural network that can enhance spatial feature extraction by emphasizing the relationship between channels with dot product-based channel attention and 3D convolution. The proposed method showed superior performance than the comparative models by achieving an accuracy of 0.58 for 4-class visual imagery (VI) EEG classification. Through statistical and neurophysiological analysis, visual motion imagery showed higher $\alpha$-power spectral density (PSD) over the visual cortex than static VI. Also, the VI of swarm dispersion showed higher $\beta$-PSD over the pre-frontal cortex than the VI of swarm aggregation.
\end{abstract}


\section{Introduction}
Brain-computer interface (BCI) is a promising technology to control the peripheral devices using intrinsic brain activity. BCI has been developed to rehabilitate neurological impairment patients\cite{lee2018high, jeong2020brain}, estimate one's mental state\cite{lee2020frontal,lee2020continuous,shin2020assessment}, and detect some diseases early \cite{zhang2017hybrid,zhang2019strength}. Electroencephalogram (EEG) is one of the most actively used non-invasive BCI modalities. EEG systems have a relatively lower equipment price than other modalities, and their signals have a relatively high temporal resolution. 

Various kinds of EEG-based BCI paradigms have been developed to induce particular brain signals in a specific condition. Exogenous BCI paradigms such as P300\cite{fazel2012p300}, event-related potential, or steady-state visual evoked potential\cite{kwak2017convolutional,liu2014recent,kwak2015lower} use visual cues to induce the user's responses. The endogenous BCI paradigms, including motor imagery (MI)\cite{Cho2021Neurograsp,kwon2019subject} and visual imagery (VI)\cite{pearson2019human,sousa2017pure}, allow users to control external devices by extracting and classifying the user intention from intrinsic brain signals\cite{suk2011subject}. However, MI is not very intuitive for people with motor disabilities, and about 20\% of the BCI users suffer from “BCI illiteracy,” which is inefficient to use the MI paradigm. On the other hand, various of VI paradigms such as static VI\cite{kosmyna2018attending}, visual motion imagery\cite{kwon2020decoding}, and mental rotation\cite{zoefel2011neurofeedback} showed the possibility of VI as an intuitive control strategy.

Recently, numerous deep learning algorithms have been designed to classify EEG signals. A convolutional neural network (CNN) has been successfully applied to EEG-based BCIs for end-to-end feature extraction and classification\cite{lee2020classification, lee2021subject}, as well as computer vision and speech recognition. Schirrmeister \textit{et al}.\cite{schirrmeister2017deep} proposed CNN architectures to decode raw EEG signals with a range of different architectures. Although the features were not fixed priors, they achieved as good performance as filter bank common spatial pattern (FBCSP)\cite{ang2008filter}, which has been most popularly applied to feature extraction for MI classification using EEG. Lawhern \textit{et al}.\cite{lawhern2018eegnet} designed a single CNN architecture that is robust enough to classify EEG signals from different BCI paradigms (P300 visual-evoked potentials, movement-related cortical potentials, error-related negativity responses, and sensory-motor rhythms).
Some studies employed 3D convolution with 3D representation of EEG to preserve the EEG spatial representations\cite{zhao2019multi,zhang2019making}. Still, most of these approaches result in sub-optimal spatial feature extraction due to null values around the corners of the channel montage. 

In this study, to classify VI-based EEG, we propose the attention-based dual-stream 3D-CNN (ADS-3D-CNN). For the data acquisition, we designed the experimental protocol to classify EEG signals depending on the VI of different swarm behaviors. We conducted statistical and neurophysiological analyses to investigate spatial-spectral features during VI. With the neurological basis, we attempted to utilize the brain connectivity during VI as a spatial-temporal feature. According to our observation, the proposed method is more reliable and accurate than other conventional methods in decoding VI EEG signals. 
 
\section {Materials and Methods}
\subsection{Participants}
Twenty healthy subjects (S1-S20, twenty males, aged 25.5 ($\pm 3.1$)) participated in the experiments. The Institutional Review Board at Korea University approved all experiments (KUIRB-2020-0318-01). Before the experiment, we informed them to get adequate sleep (over seven hours) and avoid any alcohol the day before. Subjects were informed about the experimental protocols and procedures. They provided their written consent according to the Declaration of Helsinki.

\subsection{EEG Signal Acquisition}
We measured the subjects' EEG signals using BrainAmp (BrainProduct GmbH, Germany) and 64 Ag/AgCl electrodes according to the 10/20 international system. The reference electrode was placed at FCz, and the ground electrode was placed at FPz. We set up the sampling rate to 500 Hz, and a 60 Hz notch filter was applied. Before acquiring EEG data, all electrodes' impedance was kept below 10 k$\Omega$ by injecting conductive gel.
%%%%%%%%%%%%%%%%%%%%%%%%%%%%%%%%%%%%%%%%%%%%%%%%%%%%%%%%%%%%%%%%%%%
\begin{figure}[!]
\centering
\scriptsize
\centerline{\includegraphics[width=\columnwidth]{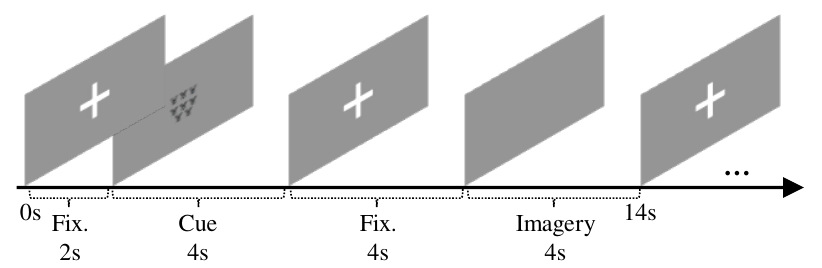}}
\caption{The overview of experimental protocol. Fixation cross (Fix.) was presented for 4s before the imagery in order to reduce the visual afterimage.}
\end{figure}
%%%%%%%%%%%%%%%%%%%%%%%%%%%%%%%%%%%%%%%%%%%%%%%%%%%%%%%%%%%%%%%%%%%
%%%%%%%%%%%%%%%%%%%%%%%%%%%%%%%%%%%%%%%%%%%%%%%%%%%%%%%%%%%%%%%%%%%
\begin{figure}[!]
\centering
\scriptsize
\centerline{\includegraphics[width=\columnwidth]{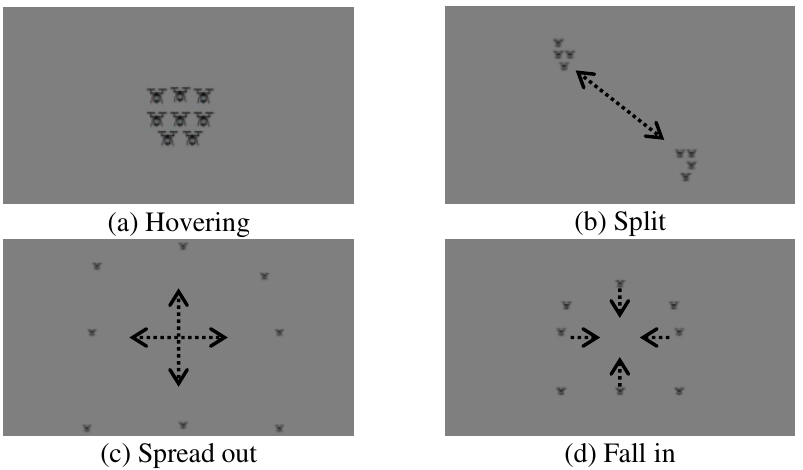}}
\caption{Detailed visual cue for the VI experiment. `Hovering' class was presented as a static image, while the rest of classes were presented as moving images.}
\end{figure}
%%%%%%%%%%%%%%%%%%%%%%%%%%%%%%%%%%%%%%%%%%%%%%%%%%%%%%%%%%%%%%%%%%%
\subsection{Experimental Procedure}
Fig. 1 shows the experimental protocol for each trial. In order to minimize the eye movement, a fixation cross was presented for 2 seconds at the start of each trial. The visual cue for each class was randomly presented for 4 seconds. After the visual stimulus, a fixation cross was presented for 4 seconds to reduce the visual afterimage. Then, a blank screen was shown for 4 seconds to imagine the previous swarm behavior visually. The whole trial took 14 s, and 50 trials were repeated for each of the four classes.

Fig. 2 indicates a detailed visual cue of the experiment. For the VI tasks, subjects were instructed to visually imagine the scene of four different swarm behaviors after watching the image of corresponding tasks (`split,' `spread out,' `fall in,' and `hovering'). In order to analyze the difference between the imagination of a static image and a moving image, the image for the `hovering' class was presented as a static image, and the rest of the images were presented as moving images. The conditions for each class are followed by the next statement.
\begin{itemize}
\item `Hovering': The static image of a particle swarm is levitating.
\item `Split': The initial position is the same as the `hovering' condition. A swarm is divided into two groups and move in opposite directions.
\item `Spread out': The initial position is the same as the `hovering' condition. A swarm disperses and directs away from the central position.
\item `Fall in': The individual particles are arranged far apart, then aggregated to the central position.
\end{itemize}

%%%%%%%%%%%%%%%%%%%%%%%%%%%%%%%%%%%%%%%%%%%%%%%%%%%%%%%%%%%%%%%%%%%%%%%%%%%%%%%%%%%%%%%
\begin{figure}[!]
\scriptsize
\centerline{\includegraphics[width=\columnwidth]{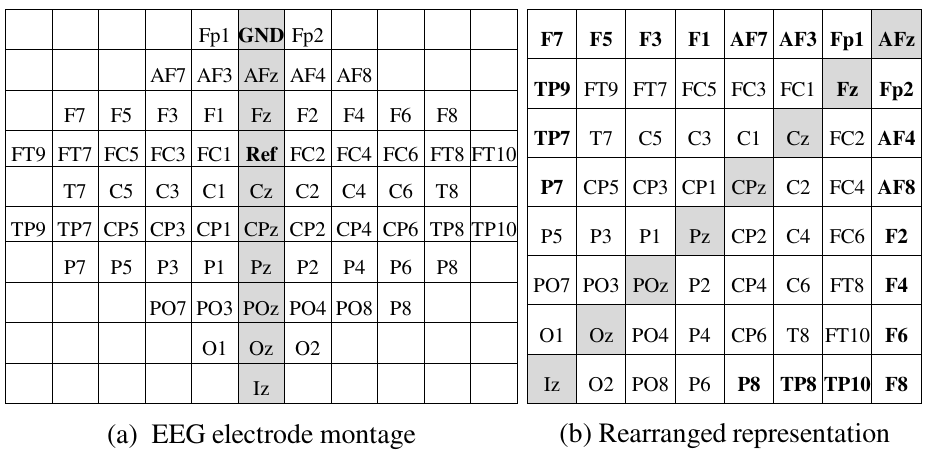}}
\caption{The EEG electrode montage (a) and rearranged representation (b). 64ch EEG electrode montage rearranged to 8$\times$8 matrix form in order to remove zeros around the corners of the channel montage.}
\end{figure}
%%%%%%%%%%%%%%%%%%%%%%%%%%%%%%%%%%%%%%%%%%%%%%%%%%%%%%%%%%%%%%%%%%%%%%%%%%%%%%%%%%%%%%%
%%%%%%%%%%%%%%%%%%%%%%%%%%%%%%%%%%%%%%%%%%%%%%%%%%%%%%%%%%%%%%%%%%%
\begin{figure*}[!]
\centering
\scriptsize
\centerline{\includegraphics[width=\textwidth]{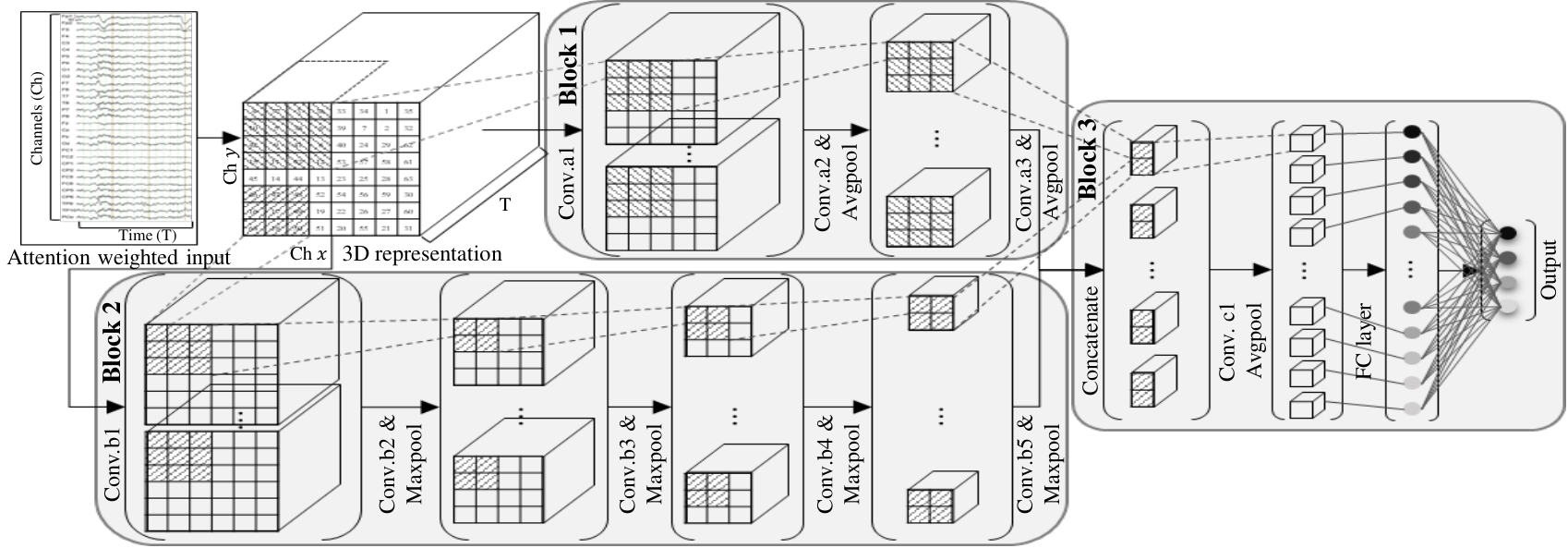}}
\caption{Overall framework of the ADS-3D-CNN. A dot-product attention was applied to EEG input and rearranged as a 3D representation. The 3D inputs proceed to two CNN pipeline, block 1 and 2. Consecutively, the output features of block 1 and 2 were concatenated and pass through another convolution block (block 3).}
\end{figure*}

%%%%%%%%%%%%%%%%%%%%%%%%%%%%%%%%%%%%%%%%%%%%%%%%%%%%%%%%%%%%%%%%%%%
\subsection{Data Analysis}
EEGLAB Toolbox\cite{delorme2004eeglab} (version 2021.0) and BBCI toolbox \cite{blankertz2010berlin} were used for data analyses. 
For EEG classification, the data were downsampled to 250 Hz and bandpass filtered with a fifth-order Butterworth filter in the frequency range of 4 to 40 Hz in order to filter electrooculogram (EOG) artifacts on $\delta$-band (0.5-4 Hz) and electromyogram artifacts at high frequency. 160 trials (40 trials $\times$ 4 classes) were used for the training dataset, and 40 trials (10 trials $\times$ 4 classes) were used for the test dataset.

To investigate the difference among VI tasks for various frequency bands, a two‐way analysis of variance (ANOVA, at a significance level of \textit{p}$<$0.05) was applied to the spectral power. One factor was the class, and the other factor was the channel (spatial information).
Three frequency bands ($\alpha$-band: 8-13 Hz, $\beta$-band: 13-30 Hz, and both frequency bands: 8-30 Hz) were selected to extract the power spectral density (PSD), respectively. The statistical comparisons of PSD were performed by paired \textit{t}‐test with Bonferroni's correction for post-hoc analysis. EEGLAB toolbox was used to plot the scalp topographies of paired \textit{t}-test results.

\subsection{Attention-based Dual-Stream 3D-CNN}
As a decoding method, we proposed the attention-based dual-stream 3D-CNN (ADS-3D-CNN). Before the EEG data were rearranged to 3D representation, scaled dot-product attention\cite{vaswani2017attention} has applied to focus the inter-channel relationship. The input data is first linearly transformed into vectors, queries (\textit{Q}) and keys (\textit{K}) of dimension $d_{k}$, and values (\textit{V}) of dimension $d_{v}$, along the spatial feature dimension. \textit{Q} represents each channel that will be used to match with \textit{K} represents all the other channels using dot product. Then the result is divided by a scaling factor of $\sqrt{d_{k}}$ to ensure that the softmax function has a good perception ability. The output weight score is assigned to \textit{V} for the final representation using dot product. The process for channel attention can be expressed as
\begin{equation}\label{eq1}
Attention(Q,K,V)=Softmax({QK^T\over \sqrt{d_k}})V
\end{equation}
where \textit{Attention(Q,K,V)} is the weighted representation. \textit{Q}, \textit{K}, \textit{V} are matrices packed by vectors for simultaneous calculation.

Fig. 3 shows the EEG electrode montage (a) and rearranged channel distribution (b). In Fig. 3, channels along the central line (gray tone) of (a) were mapped into diagonal line of (b). Fz and AFz were shifted to bridge the gap of reference position (FPz). The row elements on the left hemisphere (odd numbers) and right hemisphere (even numbers) in (a) were filled along with horizontal (left) and vertical (down) directions from the diagonal position of (b) as a baseline. Elements in a bold figure of (b) represent irregular arrangements in order to maintain an 8$\times$8 shape. We expand this 2D array to a 3D array by using EEG’s temporal sequence. 

Fig. 4 depicts the overall framework of the ADS-3D-CNN. We used 3D CNN to extract features from spatial dimension and temporal dimension simultaneously. We implement two streams of CNN blocks (block 1 and 2) with different numbers of layer, pooling methods, and kernel sizes to explore various spatial-temporal features.  Then we fusion the output features with an additional CNN block (block 3). 
% \usepackage{booktabs}
% \usepackage{multirow}
\begin{table}[]
\caption{Detailed architecture of the ADS-3D-CNN}
\Huge
\renewcommand{\arraystretch}{1.4}
\resizebox{\columnwidth}{!}
{% 
\begin{tabular}{@{}cccccc@{}}
\toprule
Block              & Layer   & Operations & Kernel size     & Stride      & Output size        \\ \midrule
0                  & Reshape &                  &                 &             & {[}\textit{B}, 1, 8, 8, 1001{]} \\ \midrule
\multirow{5}{*}{1} & Conv.a1 & Batchnorm        & {[}4, 4, 125{]} &             & {[}\textit{B}, 10, 5, 5, 877{]} \\
                   & Conv.a2 & Batchnorm, ELU    & {[}3, 3, 16{]}  &             & {[}\textit{B}, 20, 3, 3, 862{]} \\
                   & Avgpool & Dropout (0.5)     & {[}1, 1, 4{]}     & {[}1, 1, 4{]} & {[}\textit{B}, 20, 3, 3, 215{]} \\
                   & Conv.a3 & Batchnorm, ELU    & {[}3, 3, 16{]}  &             & {[}\textit{B}, 30, 1, 1, 200{]} \\
                   & Avgpool & Dropout (0.5)     & {[}1, 1, 4{]}     & {[}1, 1, 4{]} & {[}\textit{B}, 30, 1, 1, 50{]}  \\ \midrule
\multirow{9}{*}{2} & Conv.b1 & Batchnorm        & {[}3, 3, 62{]}  &             & {[}\textit{B}, 6, 6, 6, 940{]}  \\
                   & Conv.b2 & Batchnorm, ELU    & {[}3, 3, 10{]}  &             & {[}\textit{B}, 12, 4, 4, 931{]} \\
                   & Maxpool & Dropout (0.25)    & {[}1, 1, 2{]}     & {[}1, 1, 2{]} & {[}\textit{B}, 12, 4, 4, 465{]} \\
                   & Conv.b3 & Batchnorm, ELU    & {[}2, 2, 10{]}  &             & {[}\textit{B}, 18, 3, 3, 456{]} \\
                   & Maxpool & Dropout (0.25)    & {[}1, 1, 2{]}     & {[}1, 1, 2{]} & {[}\textit{B}, 18, 3, 3, 228{]} \\
                   & Conv.b4 & Batchnorm, ELU    & {[}2, 2, 10{]}  &             & {[}\textit{B}, 24, 2, 2, 219{]} \\
                   & Maxpool & Dropout (0.25)    & {[}1, 1, 2{]}     & {[}1, 1, 2{]} & {[}\textit{B}, 24, 2, 2, 109{]} \\
                   & Conv.b5 & Batchnorm, ELU    & {[}2, 2, 10{]}  &             & {[}\textit{B}, 30, 1, 1, 100{]} \\
                   & Maxpool & Dropout (0.25)    & {[}1, 1, 2{]}     & {[}1, 1, 2{]} & {[}\textit{B}, 30, 1, 1, 50{]}  \\ \midrule
\multirow{3}{*}{3} & Concatenate  &                  &                 &             & {[}\textit{B}, 30, 1, 2, 50{]}  \\
                   & Conv.c1 & Batchnorm, ELU    & {[}1, 2, 10{]}  &             & {[}\textit{B}, 60, 1, 1, 41{]}  \\
                   & Avgpool & Dropout (0.5)     & {[}1, 1, 2{]}     & {[}1, 1, 2{]} & {[}\textit{B}, 60, 1, 1, 20{]}  \\ \bottomrule
\end{tabular}}
\end{table}
%%%%%%%%%%%%%%%%%%%%%%%%%%%%%%%%%%%%%%%%%%%%%%%%%%%%%%%%%%%%%%%%%%%%

Table I shows the detailed architecture of our proposed ADS-3D-CNN. 
The batch size (\textit{B}) is set to 40. The CNN block 1 consists of three convolution layers and two average pooling layers. The CNN block 2 consists of five convolution layers and four max-pooling layers. The size of the temporal kernel of conv.a1 and conv.b1 are set to half and quarter of the sampling rate, to capture frequency information above 2 Hz and 4 Hz, repectively.
For the feature fusion, blocks 1 and 2 were designed to have the same output size in the last layers. The output features were concatenated to the channel dimension, then block 3 fusion the two features.

In order to resolve the covariate shift problem, batch normalization was used after each convolution layer. Then, pooling layers were set to resize the convolution. Exponential linear unit (ELU) was used as an activation function. We used optimizer AdamW\cite{loshchilov2017decoupled} to minimize the categorical cross-entropy loss function. We performed 200 iterations for the model training process and saved the model weights and hyper-parameters that produced the lowest loss of the test data.

%%%%%%%%%%%%%%%%%%%%%%%%%%%%%%%%%%%
\begin{table}[!]
\caption{Decoding performance comparison with the FBCSP, DeepConvNet, ShallowConvNet, EEGNet, and the proposed method}
\Large
\renewcommand{\arraystretch}{1.1}
\resizebox{\columnwidth}{!}
{% 
\begin{tabular}{@{}cccccc@{}}
\toprule
\multirow{3}{*}{Subjects} & \multicolumn{5}{c}{Methods}                   \\ \cmidrule(l){2-6} 
                          & \begin{tabular}[c]{@{}c@{}}FBCSP\\ \cite{ang2008filter}\end{tabular}& \begin{tabular}[c]{@{}c@{}}DeepConvNet\\ \cite{schirrmeister2017deep}\end{tabular}& \begin{tabular}[c]{@{}c@{}}ShallowConvNet\\ \cite{schirrmeister2017deep}\end{tabular} & \begin{tabular}[c]{@{}c@{}}EEGNet\\ \cite{lawhern2018eegnet}\end{tabular}& Proposed \\ \midrule
S1                        & 0.38  & 0.58 & 0.45    & 0.55   & 0.60        \\
S2                        & 0.38  & 0.55 & 0.45    & 0.50   & 0.55        \\
S3                        & 0.38  & 0.35 & 0.53    & 0.60   & 0.65        \\
S4                        & 0.30  & 0.38 & 0.55    & 0.55   & 0.55        \\
S5                        & 0.30  & 0.53 & 0.50    & 0.50   & 0.53        \\
S6                        & 0.38  & 0.45 & 0.55    & 0.45   & 0.60        \\
S7                        & 0.30  & 0.55 & 0.53    & 0.53   & 0.63        \\
S8                        & 0.33  & 0.53 & 0.60    & 0.65   & 0.68        \\
S9                        & 0.30  & 0.43 & 0.65    & 0.68   & 0.70        \\
S10                       & 0.30  & 0.45 & 0.45    & 0.58   & 0.55        \\
S11                       & 0.28  & 0.38 & 0.48    & 0.58   & 0.53        \\
S12                       & 0.33  & 0.58 & 0.50    & 0.53   & 0.53        \\
S13                       & 0.30  & 0.43 & 0.53    & 0.55   & 0.63        \\
S14                       & 0.33  & 0.48 & 0.53    & 0.55   & 0.53        \\
S15                       & 0.33  & 0.43 & 0.53    & 0.58   & 0.55        \\
S16                       & 0.35  & 0.53 & 0.58    & 0.48   & 0.63        \\
S17                       & 0.28  & 0.55 & 0.55    & 0.58   & 0.55        \\
S18                       & 0.28  & 0.53 & 0.53    & 0.48   & 0.50        \\
S19                       & 0.30  & 0.43 & 0.53    & 0.58   & 0.63        \\
S20                       & 0.33  & 0.45 & 0.50    & 0.48   & 0.55        \\ \hline
Average                   & 0.32  & 0.48 & 0.53    & 0.55   & \textbf{0.58}        \\ \hline
Std.                      & 0.03  & 0.07 & 0.05    & 0.06   & \textbf{0.05}        \\ \hline
\end{tabular}}
\end{table}
%%%%%%%%%%%%%%%%%%%%%%%%%%%%%%%%%%%

\section{Results and Discussion}
We compare the decoding performance of our proposed ADS-3D-CNN with a traditional machine learning algorithm (FBCSP\cite{ang2008filter}) and the conventional CNNs (DeepConvNet\cite{schirrmeister2017deep}, ShallowConvNet\cite{schirrmeister2017deep}, and EEGNet\cite{lawhern2018eegnet}). Within-subject five-fold cross-validation results across all algorithms for 4-class VI EEG classification are shown in Table II. The average accuracy of the proposed method was 0.58, which outperforms the conventional methods. The results indicate that the average classification accuracy of the ADS-3D-CNN was improved by 0.03 and 0.05 compared to that of EEGNet and ShallowConvNet, respectively. In particular, S9 showed the highest classification accuracy of 0.70. Moreover, the proposed method has a lower standard deviation for classification accuracies among the subjects, which suggests that our proposed method is more reliable to decode EEG signals than other conventional methods. 
Channel-wise attention demonstrates superior performance because it can change the weight of different channels to explore the information of a feature map. The 3D-convolution method and the dual-stream CNN can preserve the temporal features of the EEG and the spatial features of the brain region, enhance the resolution and analyze the EEG signals from different views. The overall results suggest that the attention-based dual-stream 3D-CNN framework can improve the performance of VI EEG classification.

%%%%%%%%%%%%%%%%%%%%%%%%%%%%%%%%%%%%%%%%%%%%%%%%%%%%%%%%%%%%%%%%%%%%%%%%%%%%%%%%%%%%%%%
\begin{figure}[!]
\scriptsize
\centerline{\includegraphics[width=\columnwidth]{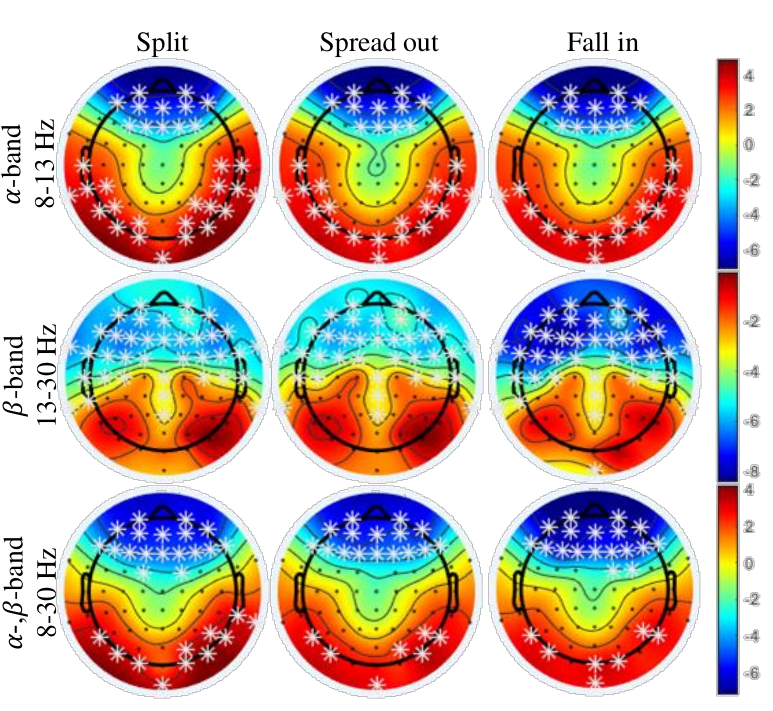}}
\caption{Differences in spectral power between different classes and frequency ranges. The statistical results represent \textit{t}-values in each frequency band for differences between the visual motion imageries (`split,' `spread out,' and `fall in') and static VI (`hovering'). The white asterisk indicates a significant electrode in spectral power (\textit{p}$<$0.01 with Bonferroni's correction).}
\end{figure}
%%%%%%%%%%%%%%%%%%%%%%%%%%%%%%%%%%%%%%%%%%%%%%%%%%%%%%%%%%%%%%%%%%%%%%%%%%%%%%%%%%%%%%%
To examine the significant spatial and spectral features for decoding VI, we calculated the spectral power in three frequency bands ($\alpha$-, $\beta$-, and both bands). Here, we focused on $\alpha$- and $\beta$-bands since the $\alpha$- and $\beta$-activities are related to the visual information process\cite{xie2020visual} and semantic processing\cite{hamame2012reading}, respectively. 

Fig. 5 represents the spatial differences in spectral power between different classes of VI task and frequency ranges using the Bonferroni corrected paired \textit{t}-test. The statistical results represent \textit{t}-values in each frequency band for differences between the visual motion imageries (`split,' `spread out,' and `fall in') and static VI (`hovering'). The blue color reflects relatively lower activity, whereas the red color reflects higher activity compared to another class. The white asterisk indicates a significant electrode in spectral power (\textit{p}$<$0.01 with Bonferroni's correction).
Across all visual motion imagery tasks, the temporal-occipital region had higher $\alpha$-PSD, indicating that higher intensity of brain activation was represented on the visual cortex during visual motion imagery than static VI. Also, the pre-frontal region had lower PSD during visual motion imagery tasks than static VI. 
$\beta$-PSD showed different representations in the pre-frontal region among the classes. 
Specifically, the VI of swarm dispersion (`split' and `spread out') showed higher PSD over $\beta$-band in the pre-frontal cortex than the VI of swarm aggregation (`fall in'), while there were no statistically significant channels on occipital region. 
In the case of $\alpha$- and $\beta$-bands (8-30 Hz), although the overall representations were similar to the case of $\alpha$-band, the PSD values of the right occipital region were stronger during the `split' task compared to the `fall in' task, while the overall temporal PSD values were decreased.

\section{Conclusion and Future works}
In this work, we proposed attention-based dual-stream 3D-CNN to classify VI tasks. Experimental results show that our proposed ADS-3D-CNN achieved superior performance in VI classification with rearranged 3D representation. We verified our experimental VI paradigm through the neurophysiological and statistical analysis and found the significant channels depending on VI tasks and spectral bands. For future work, additional high-density EEG recording and EEG source localization is needed to investigate intra-cortical activity during VI.

\bibliographystyle{IEEEtran}
\bibliography{ref}
\end{document}